\documentclass{emulateapj}
\usepackage{epsfig,graphicx}

 \renewcommand\appendix{\par 
   \setcounter{figure}{0}%
   \renewcommand\thefigure{A.\arabic{figure}}} 

\slugcomment{Accepted by The Astrophysical Journal Letters on October
  15, 2009}

\shorttitle{The First Generation of Virgo dEs?}
\shortauthors{T. Lisker et al.}

\begin{document}
 
\title{The First Generation of Virgo Cluster Dwarf Elliptical Galaxies?}

\author{Thorsten Lisker\altaffilmark{1}, Joachim Janz\altaffilmark{1,2,8},
  Gerhard Hensler\altaffilmark{3}, Suk Kim\altaffilmark{4}, Soo-Chang
  Rey\altaffilmark{4}, Simone Weinmann\altaffilmark{5}, Chiara Mastropietro\altaffilmark{6}, Oliver
  Hielscher\altaffilmark{1}, Sanjaya
  Paudel\altaffilmark{1}, Ralf Kotulla\altaffilmark{7}
  }
\affil{
$^1$ Astronomisches Rechen-Institut, Zentrum f\"ur Astronomie der
  Universit\"at Heidelberg, M\"onchhofstra\ss e 12-14, 69120
  Heidelberg, Germany; TL@x-astro.net\\
$^2$ Division of Astronomy, Department of Physical Sciences,
  University of Oulu, P.O.\ Box 3000, FIN-90014 Oulu, Finland\\
$^3$ Institute of Astronomy, University of Vienna, T\"urkenschanzstrasse 17, 1180 Vienna, Austria\\
$^4$ Department of Astronomy and Space Science, Chungnam National University, Daejeon 305-764, Korea\\
$^5$ Max-Planck-Institut f\"ur Astrophysik, Karl-Schwarzschild-Stra\ss e 1, 85748 Garching, Germany\\
$^6$ LERMA, Observatoire de Paris, UPMC, CNRS, 61 Av.\ de l'Observatoire, 75014 Paris, France\\
$^7$ Centre for Astrophysics Research, University of Hertfordshire,
College Lane, Hatfield AL10 9AB, UK\\
}
\altaffiltext{8}{Fellow of the Gottlieb Daimler and Karl Benz Foundation}

\begin{abstract}
In the light of the question whether most early-type dwarf (dE) galaxies in
clusters formed through infall and transformation of late-type
progenitors, we search for an imprint of such an infall history in
the oldest, most centrally concentrated dE subclass of the Virgo
cluster: the nucleated dEs that show no signatures of disks or
 central residual star formation.
We select dEs in a (projected) region around the central elliptical
galaxies, and subdivide them by their line-of-sight velocity into
fast-moving and slow-moving ones. These subsamples turn out to have significantly
different shapes: while the fast dEs are relatively flat objects, the
slow dEs are nearly round. Likewise, when subdividing the central dEs
by their projected axial ratio into flat and round ones, their
distributions of line-of-sight velocities differ significantly: the
flat dEs have a broad, possibly two-peaked distribution, whereas the
round dEs show a narrow single peak.
 We conclude that the round dEs probably are on circularized
  orbits, while the flat dEs are still on more eccentric or radial
  orbits typical for an infalling population. In this picture, the
  round dEs would have resided in the cluster already for a long time, or
  would even be a cluster-born species, explaining their nearly circular
  orbits. They would thus be the first generation of Virgo cluster
  dEs. Their shape could be caused by dynamical heating through
  repeated tidal interactions. Further investigations through stellar
  population measurements and studies of simulated galaxy clusters
  would be desirable to obtain definite conclusions on their origin.
\end{abstract}
 
 

\section{Introduction: the variety of early-type dwarfs}
 \label{sec:intro}

Early-type dwarf galaxies largely outnumber all other galaxy types in
rich galaxy clusters \citep[e.g.,][]{vcc}, while they are less abundant
on average in galaxy groups, and almost absent in the field
\citep{gu06}. Among dwarf galaxies, the fraction of those with early-type
morphology is significantly larger in dynamically more evolved
environments \citep{trentham02}. Apart from this global trend,
early-type dwarfs also show a
pronounced relation with local galaxy density within a given
cluster or group: \citet{bin87} found that their number strongly
increases with local density, analogous to the correlation for
giants \citep{dre80}, and \citet{tully08} observed that early-type
dwarfs strongly cluster around major E/S0 galaxies.
These findings have often been interpreted as indicating that
early-type dwarfs are actually formed through environmental processes,
thereby transforming late-type into early-type objects. Such
mechanisms, mainly gas stripping and tidal effects on structure and
star formation, have been investigated in various studies
\citep[e.g.,][]{moo98,vZe04a,sab05,boselli08}, with the
main conclusion that one or more of these mechanisms must be
responsible for the majority of cluster dEs.

Despite the apparently simple structure of early-type dwarfs, a
closer investigation of their characteristics revealed a surprising
complexity. Disk signatures, such as bars and weak spiral arms \citep[cf.][]{jer00a}, were
identified in a significant fraction of bright Virgo cluster dEs
\citep[and references therein]{p1}, confirming earlier indications of disks \citep{san84,bin91}.
The distribution of projected axial ratios reveals that these objects
are not spheroidal, but are rather shaped like thick disk galaxies
\citep{p3}. Similar shapes, though somewhat thicker, are found for the
brighter nonnucleated dEs \citep{fer89}, as well as for those with
blue central colors from ongoing or very recent residual star
formation \citep{vig84,p2}.
These dE subclasses do not show strong clustering,
but are distributed more like the late-type cluster galaxies
\citep{p3}, lending support to the above formation
scenarios.

\begin{figure*}
  \epsscale{1.0}
  \plotone{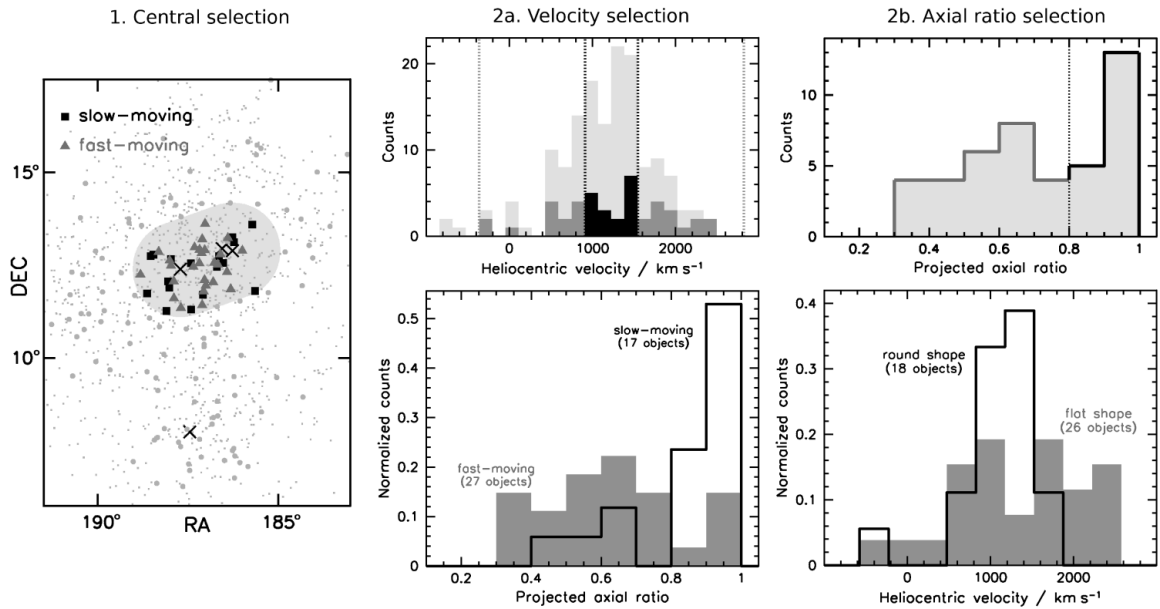}
  \caption{{\bf Selection process and resulting subsamples.}
    \emph{Left:} Positions of
    all Virgo cluster member galaxies (grey dots), of those 149 dE(N)s
    for which heliocentric velocities are available (grey circles),
    and of selected fast-moving and slow-moving dE(N)s (grey triangles
    and black squares, respectively; see the middle panel). The
    positions of the major 
    elliptical galaxies are indicated by crosses (M49, M84, M86, M87,
    counterclockwise from South). The central region (see text) is
    shown as grey-shaded area.
    \emph{Middle, top:}
    Distribution of
    heliocentric velocities of all dE(N)s (light grey), central
    fast-moving dE(N)s 
    (dark grey), and central slow-moving dE(N)s (black). The black
    vertical lines mark the semi-interquartile range ($\pm 320$ km/s)
    of velocities relative to the median
    ($1228$ km/s), calculated from all dE(N)s. The grey
    vertical lines mark five semi-interquartile ranges, beyond which
    objects are excluded.
    \emph{Middle, bottom:} 
    Distribution of projected axial ratios for the fast and slow subsamples.
    \emph{Right, top:}
    Distribution of projected axial ratios of central
    dE(N)s. We subdivide them at
    an axial ratio of 0.8 into a flat and a round
    subsample.
    \emph{Right, bottom:} Resulting distribution of heliocentric
    velocities for the two subsamples.
  }
  \label{fig:select}
\end{figure*}

Only the nucleated dEs that do not show disk features or
blue cores appear to keep up the image of dEs as spheroidal objects
consisting of old stars, concentrated in regions of high
local density \citep{fer89,p3}. They are the dE subclass with the
oldest stellar populations \citep{p4,Paudel2009},
and were concluded to be potential primordial objects in a study of
the Coma and Fornax clusters \citep{rak04}. Indeed, semi-analytic
galaxy evolution models tied to N-body simulations
of hierarchical structure formation show at least qualitative
consistency with the observed scaling relations and colors of early-type dwarf and
giant galaxies \citep{DeRijcke2005,janz08,janz09a}, indicating that a
``cosmological'' formation of dEs from 3-sigma density peaks on top of the
cluster potential is plausible as well.
In the study presented here, our
goal is to assess whether even the centrally concentrated nucleated dEs
of the Virgo cluster show some indication of an infall history,
linking their formation process to environmental effects.

\section{Sample selection and data}
 \label{sec:data}

As central cluster region, we select an area encompassing the
massive elliptical galaxies M84, M86, and M87, requiring that the
projected distance to M87, to M84, or to their connecting axis, is less
than 1.27$\arcdeg$, corresponding to 0.35 Mpc with $m-M=31$ mag
(Fig.~\ref{fig:select}, left). This selection 
accounts for the fact that the dEs are not only clustered around M87,
which marks the intracluster gas mass center \citep{boehringer94},
but their distribution peaks in the region between M84/M86 and M87
\citep{bin87}.

Based on the study of dE subclasses by \citet{p3}, we select only
nucleated dEs (\,dE(N)s\,) that display no disk
features or blue central colors \citep[cf.][]{p1,p2}. Only certain
cluster members \citep{vcc,virgokin}
brighter than $m_\mathrm{B}\le18.0$ mag are taken into account. This is the
same magnitude limit up to which the Virgo cluster catalog
\citep[VCC,][]{vcc} was found to be complete. 
Our classification of nucleated and non-nucleated
dEs relies on the VCC.
However, many apparently non-nucleated dEs actually host a
faint nucleus hardly detectable with the imaging data on which the VCC was
based \citep{gra05,acsvcs8}. Nevertheless, using the VCC
classification yielded the said differences in structure, color, and
distribution of the subclasses, which is why we stick to this definition
\citep[also see the discussion in][]{p3}.

Heliocentric velocities are available for 46 dE(N)s in the central region
through the
NASA/IPAC Extragalactic Database (NED), of which 29 are provided by the Sloan
Digital Sky Survey (SDSS) data release 4 \citep{sdssdr4}, and 17 from
other sources
\citep{vcc,virgokin,conI,2002A&A...384..371S,evstigneeva07}.
Outer axial ratios are measured by fitting an ellipse to the
$r$-band isophote at a semimajor axis $a = 2 a_{{\rm hl},r}$, where
$a_{{\rm hl},r}$ defines the half-light aperture \citep{p3}.
Ultraviolet-optical colors (Kim et al.\ 2009, in preparation) are obtained by
combining total far-ultraviolet ($FUV$) and near-ultraviolet ($NUV$) magnitudes provided by GALEX
\citep[Galaxy Evolution Explorer,][]{galex} with total optical magnitudes, measured by \citet{p4} on
SDSS data release 5 images \citep{sdssdr5}.

\section{Result: galaxy shape versus velocity}
 \label{sec:result}

From the heliocentric velocities, we calculate relative
line-of-sight
 velocities,
using the
median value $\eta = 1228$ km/s of all\footnote{Velocities are available through NED for 149
  of 210 dE(N)s that are certain cluster members.} Virgo cluster dE(N)s as zeropoint.
We then use the semi-interquartile range $\xi=320$ km/s of all dE(N)s
to divide the central dE(N)s into fast-moving and slow-moving
ones
with respect to the line of sight,
according to their relative velocities
(Fig.~\ref{fig:select}, top middle), not distinguishing between
the direction of movement (positive or negative velocities). Two galaxies
with velocities beyond $5\xi$ are excluded, leaving 27 fast and 17 slow
dE(N)s.

The projected axial ratios of fast and slow galaxies turn out to be
significantly different (Fig.~\ref{fig:select}, bottom middle): while the slow
ones have nearly round shapes, the fast ones are 
much more flattened. The median
axial ratios are 0.90 and 0.64 for slow 
and fast galaxies, and the mean values are 0.84 and 0.63,
respectively.
A
K-S test yields a probability of 0.1\% that both subsamples are drawn from
the same parent distribution.

The existence of two distinct subgroups is confirmed when the
selection process is reversed: we now divide the central dE(N)s based
on their projected axial ratio (Fig.~\ref{fig:select}, top right), and compare the resulting
distributions of heliocentric velocities. Again, the two subsamples are clearly
different. The round galaxies have a narrow, centrally peaked
velocity distribution, while the flatter galaxies have a much broader
distribution, possibly with a two-peak structure
(Fig.~\ref{fig:select}, bottom right).
The velocity dispersion, calculated as
standard deviation with one clipping at $\pm 2.3 \sigma$ ($\pm 1\%$ for
a normal distribution), is 737 km/s for the flatter
subsample, and only 373 km/s for the round
subsample. A \mbox{K-S} test on the distribution
of velocities relative to the median, considering only absolute values,
yields a probability of 0.1\%.

\section{Discussion}
 \label{sec:discuss}

\subsection{Subpopulations and their distribution}

\begin{figure}
  \epsscale{1.0}
  \plotone{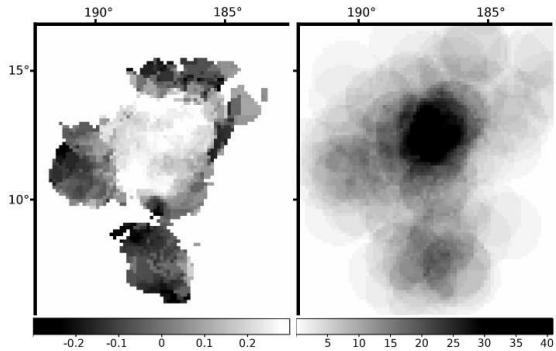}
  \caption{{\bf Map of axial ratio differences.} \emph{Left:} Map of
    the Virgo cluster, with the greyscale level indicating the
    difference between the local median axial ratio of slow and fast
    galaxies, according to the scale given below the figure. In bright
    regions, slow galaxies are rounder than fast ones, while this is
    reversed in dark regions. The map was calculated using circular
    areas of 1.27$\arcdeg$ radius (corresponding to 0.35 Mpc with $m-M=31$
    mag), sampled every 0.127$\arcdeg$ in right ascension and
    declination. This strong oversampling leads to a smooth
    map. The velocity separation is done as in Fig.~\ref{fig:select}
    (middle). A value is calculated only if there are at least three
    of each of slow and fast galaxies. \emph{Right:} Comparison map
    indicating how many galaxies are included in the local circular
    areas for which the values in the left panel were calculated.
  }
  \label{fig:map}
\end{figure}

Following \citet{conI}, the comparably high velocity dispersion of the flatter
dE(N)s (737 km/s) could be seen as typical for a galaxy population with an infall 
history; spiral and irregular galaxies have similar dispersions (776
and 727 km/s, respectively, \citealt{conI}). However, there is one caveat
to this straightforward interpretation: our
galaxies were selected to be in the (projected) central cluster
region, and there, the velocity dispersion of \emph{all} galaxy types
was found to increase \citep{conI}. The conclusion of \citeauthor{conI}
was that most galaxies are on radial or highly eccentric orbits. In
the light of these observations, the low velocity dispersion of the round
central dE(N)s is even more remarkable (373 km/s, as compared to the 462 km/s of giant
ellipticals, \citealt{conI}).
Due to the selection of galaxies in a
projected central region, it is the most plausible explanation that
those dE(N)s moving slow along the line of sight are on nearly
circular orbits, whereas most of those with high line-of-sight
velocities should be on more eccentric orbits.

If this was the case,
we should see, to some extent, a reverse effect when selecting
galaxies in the projected outskirts: those on more circular orbits
should now, on average, be the faster-moving ones along the line of
sight. Indeed, when defining ``slow'' and ``fast'' locally,
the clear axial ratio difference seen in
the central area does not only disappear towards the outer regions,
but is somewhat reversed (Fig.~\ref{fig:map}), supporting our 
interpretation.
The circularized orbits of the round, slow dE(N)s
might thus indicate that they are the first generation of dEs in the
Virgo cluster \citep[see also][]{Biviano2009}.
Based on simulated cluster galaxies \citep{springel05} in the semi-analytic
  model of \citet{delucia07}, just above the mass range of dwarfs, we find
  indications that for galaxies in the projected cluster center,
  slow-moving galaxies have typically fallen in earlier than faster
  ones.

 While those dE(N)s could represent  the low-mass
continuation of elliptical galaxies in the framework of $\Lambda$CDM
structure formation \citep{janz08,janz09a}, thus being ``primordial''
dEs \citep[cf.][]{rak04},
one could also imagine that they were
formed at early epochs out of other galaxies by
transformation processes.
On the one hand, the observations appear to be partially consistent with the galaxy
harassment scenario \citep[e.g.,][]{moo96}, in which an
infalling disk galaxy experiences several close encounters with
massive substructures and is eventually transformed into a dE.
In 
N-body simulations of harassment \citep{mas05},
the overall shape of harassed objects becomes rounder with each
further encounter. The rounder galaxies could thus have experienced
infall and transformation at earlier epochs, having resided in the
cluster for a longer time, leading to more tidal interactions.
On the other hand, a possible inconsistency appears when considering
that the strongest harassment effects occur when
galaxies are orbiting within the cluster center, causing
significant dynamical heating \citep{mas05}. This would imply that the
rounder galaxies should be orbiting in the
(threedimensional) central region, deep within the Virgo cluster
potential well. However, from an analysis of a galaxy cluster in the
Millennium-II simulation \citep{BoylanKolchin2009} with a mass
comparable to Virgo, we find no significant difference between the
clustercentric radii of galaxies moving slow and fast along the
line-of-sight.\footnote{Unfortunately, good-quality individual distances from
  surface brightness fluctuations \citep{mei07} are only available for
  3 slow and 2 fast galaxies.}
Moreover, rounder galaxies end up with smaller sizes on
average in the simulations, starting from the same progenitor mass. We do not
observe such a difference in size; only the \emph{scatter} is
somewhat larger for the rounder/slower than for the flatter/faster ones.

\subsection{Stellar population aspects}

\begin{figure}
  \plotone{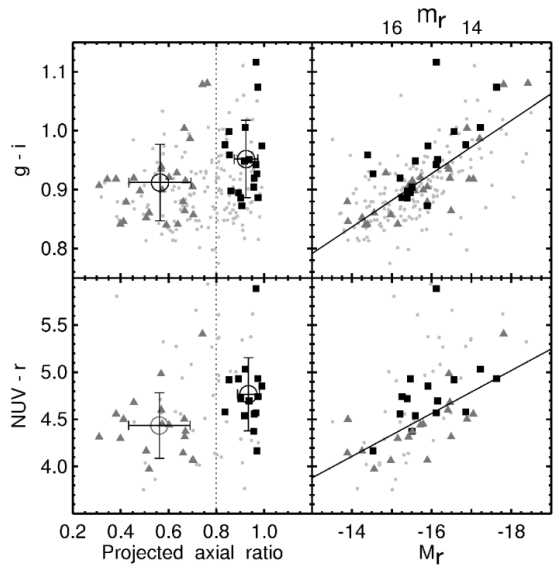}
  \caption{{\bf Ultraviolet-optical colors versus axial ratio.}
\emph{Top left:} Projected axial ratio versus $g-i$ color (measured
within the half-light aperture, \citealt{p4}) for all dE(N)s (grey
dots), the flat subsample (grey triangles), and the round subsample
(black squares). The vertical dotted line denotes the subsample
division at an axial ratio of 0.8. The circle and error bars denote
the mean and standard deviation for each subsample. \emph{Top right:}
Color-magnitude relation. The solid grey line corresponds to the
linear fit for the high-density dE(N) subsample of
\citet{p4}. \emph{Bottom panels:} Same as top panels, but for $NUV-r$
color, using total magnitudes for color calculation. Only objects
detected by GALEX in $NUV$ are shown. For absolute magnitudes we adopt
$m-M=31$ mag.
  }
  \label{fig:color}
\end{figure}

If the flatter dE(N)s experienced more recent infall, their
star formation would have been truncated later, which should be
reflected in their stellar population characteristics. Indeed, the
$g-i$ colors of the round dE(N)s are redder than those of the flat ones
(Fig.~\ref{fig:color}, top left). This is confirmed when examining the
color-magnitude relation (CMR, top right).
The difference is even clearer when the age-sensitive $NUV-r$ color is
used instead (Fig.~\ref{fig:color}, bottom).
A K-S test on the
color residuals about the CMR yields a probability of 5.0\% in $g-i$
and 2.1\% in $NUV-r$ for the same parent distribution of the two
subsamples. The values change to 7.4\% and 3.7\% when the reddest
galaxy, which is an outlier to the CMR, is omitted. 
In $FUV-r$ (not shown),
where useful measurements are only available for $\sim 50\%$ due to
detection limits, the colors of both subsamples show no difference
within the errors. 

 This behavior would be consistent
with a small ($\sim$1\,Gyr)
   age difference
 between two populations that are already relatively old: as
 illustrated in \citet{liskerhan08}, evolutionary synthesis models
 show that an aging stellar population continues to
 become redder with time in $NUV-r$, while it reaches a more or less
 constant $FUV-r$ value. In fact, when fitting SDSS $ugriz$ magnitudes, partly
 combined with near-infrared $H$ band photometry \citep[with improved
 reduction by T.\ Lisker]{goldmine}, to GALEV evolutionary
 synthesis models \citep{galev}, more than 60\% of round/slow dE(N)s
 are fit best with an exponentially declining model with $\tau=0.5$
 Gyr.\footnote{
 The model grid comprised exponentially declining models with various
 decay times $\tau$ (0.5, 1, 2, and 4\,Gyr), as well as models with
 constant star formation rate until an age $\delta$ of 2 or 8\,Gyr,
 followed by an exponential decline with $\tau$ of 1 or
 3\,Gyr. Models with fixed metallicity ($[Fe/H]$ of 0.0, -0.3, -0.7 or
 -1.7) as well as chemically consistent models were calculated. The
 fits were performed by the dedicated $\chi^2$ minimization routine
 GAZELLE written by R.\ Kotulla (in preparation).
}
 In contrast, the best-fit model of more than 60\% of
 flat/fast dE(N)s has a constant star formation rate for $\ge 2$ Gyr,
 followed by an exponential decline with $\tau\ge1$ Gyr
 \citep{hielscherthesis}.\footnote{
A significantly different mean
     age of the two subsamples is, however, not found.
Useful
     spectroscopic data are only available for 1 slow and 3 fast
     dE(N)s; a Lick index analysis yields an age of 3 Gyr for one fast
     dE(N), and ages above 10\,Gyr for the remaining objects
     \citep{Paudel2009}.}

So far, we neglected the fact that the relation between axial ratio
and color is not only seen for our sample of central dE(N)s, but is
actually present, to some extent, for the whole population of Virgo dE(N)s
(Fig.~\ref{fig:color}, grey dots). This is not a new result: \citet{bar02}
reported this very correlation, and concluded that it is a metallicity
effect. Since outflows of metal-enriched gas occur primarily along the
minor axis, flatter galaxies have a lower metallicity.
Nevertheless, the star-formation history and starburst
episodes determine the gas and metallicity loss, respectively,
from  dwarf galaxies \citep{recchi06,Valcke2008}.
It therefore remains unclear whether these two
findings, the correlation between shape and velocity on the one hand
and between shape and color on the other hand, are at all related, or
whether the latter correlation is simply inherent to all dEs,
regardless of their velocity characteristics.
More high-quality spectroscopic data are needed to investigate whether
differences in age, metallicity, and/or other stellar population
characteristics are the primary cause of the observed color
differences.

As an aside, we note that a relation might exist between the nucleus
``strength'' (i.e., its relative light contribution) and the host
galaxy's projected axial ratio: the strongest nuclei were found by
\citet{bin00} to reside in nearly round dE(N)s. However, a similar
study by \citet{gra05}, now using S\'ersic instead of King profiles
for the host galaxies, did not confirm a significant correlation
except for a mild tendency.

\section{Conclusions}
 \label{sec:conclusion}
We have established that different populations, or generations, of
Virgo cluster dE galaxies can be distinguished based on their
projected shapes and line-of-sight velocities. This is most likely
interpreted with the flatter objects moving on radial or highly
eccentric orbits, indicating an infalling galaxy population. In
contrast, the round dEs have circularized orbits, characteristic for
cluster-born galaxies or very early infall. They might thus be the
first generation of Virgo cluster dEs. While more detailed stellar
population measurements would be necessary for an unambiguous
interpretation, the observational findings themselves constitute an
important step towards understanding this most abundant
galaxy population in clusters. Our results demonstrate the
necessity to perform a similar deconvolution of dE
populations in any ongoing and future study of dEs, if we aim at
tracing back their evolutionary history along with that of their host
clusters.

\acknowledgements
    We thank the referee for constructive suggestions, Werner Zeilinger
    and Jay Gallagher for stimulating 
    discussions, Bruno Binggeli for useful comments on the
    manuscript, Marco Longhitano for kind help with software, as well
    as Theresa Gotthart and Mathias Jaeger for an
    examination of SDSS spectra.
    T.L., J.J., and S.P.\ are supported within the framework of the Excellence
    Initiative 
    by the German Research Foundation (DFG) through the Heidelberg
    Graduate School of Fundamental Physics (grant number GSC 129/1).
    J.J.\ acknowledges support by the Gottlieb Daimler and Karl Benz Foundation.
    G.H.\ acknowledges support by the FWF project P21097-N16.
    S.-C.R.\ acknowledges support from the National Research Foundation of 
    Korea (NRF) grant funded by the Korea government (MEST) (No.\ 2009-0062863).
    This study is based on the Sloan Digital Sky Survey
    (http://www.sdss.org), and
    has made use of NASA's Astrophysics Data
    System Bibliographic Services and the NASA/IPAC Extragalactic
    Database (NED).

\lastpagefootnotes


\appendix

\begin{figure}[ht]
  \epsscale{0.7}
  \plotone{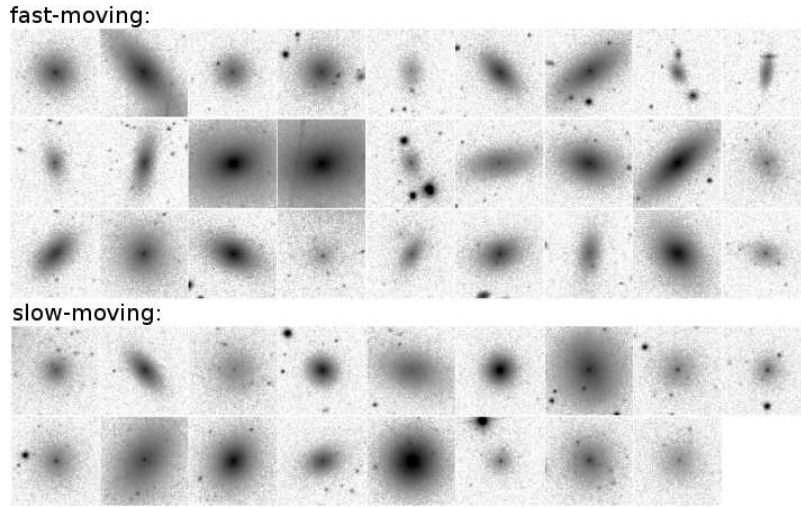}
  \caption{{\bf SDSS images of fast and slow dE(N)s.} Illustration of
    the selection by velocity shown in Fig.~\ref{fig:select}.
    The images were created by co-adding the SDSS $g$, $r$, and $i$
    bands. Contrast and scale is the same for all images. The angular
    scale is $1'\times1'$, corresponding to $4.6\times4.6$ kpc$^2$
    at $m-M=31$ mag.
  }
  \label{fig:pics}
\end{figure}

\end{document}